\documentclass[aps,prb,showpacs,amsmath,amssymb,twocolumn,superscriptaddress]{revtex4-1}
\pdfoutput=1

\usepackage{graphicx}% Include figure files
\usepackage{color}
\usepackage{amsmath}

\DeclareMathOperator{\sgn}{sgn}
\DeclareMathOperator{\Tr}{Tr}
\DeclareMathOperator{\Var}{var}
\renewcommand{\d}{\mathrm{d}}
\newcommand{\ii}{\mathrm{i}}

\renewcommand{\vec}[1]{\mathbf{#1}}
\newcommand{\mr}[1]{\mathrm{#1}}
\newcommand{\refcite}[1]{Ref.~\onlinecite{#1}}

\begin{document}

\title{Theory of temperature fluctuation statistics in superconductor-normal metal tunnel structures}

\author{M.~A.~Laakso}
\email[]{matti.laakso@aalto.fi}
\author{T.~T.~Heikkil\"a}
\affiliation{O.~V.~Lounasmaa Laboratory, Aalto University, Post Office Box 15100, FI-00076 AALTO, Finland}
\author{Yuli V.~Nazarov}
\affiliation{Kavli Institute of Nanoscience, Delft University of Technology, 2628 CJ Delft, The Netherlands}
\date{\today}

\begin{abstract}
We describe the statistics of temperature fluctuations in a SINIS structure, where a normal metal island (N) is coupled by tunnel junctions (I) to two superconducting leads (S). We specify conditions under which this structure exhibits manifestly non-Gaussian fluctuations of temperature. We consider both the Gaussian and non-Gaussian regimes of these fluctuations, and the current fluctuations that are caused by the fluctuating temperature. We also describe a measurement setup that could be used to observe the temperature fluctuations.
\end{abstract}


\maketitle

\section{Introduction}

Nanoelectronic structures with superconductor--insulator--normal metal (SIN) junctions are often used in applications involving thermometry or cooling,\cite{giazotto06} enabled by the existence of an energy gap of $2\Delta$ in the density of states of the superconductor. In those systems, typically two such junctions are connected in series to form a SINIS structure. Thermometry with such structures relies on the temperature dependence of the electric current through the SIN junction,\cite{rowell76} whereas refrigeration is based on the heat current carried by electrons from the normal metallic part.\cite{nahum94,leivo96}

In most SINIS devices the electron--electron relaxation rate is fast compared to the electron injection rate due to driving, and therefore the electron system in the device maintains its equilibrium energy distribution, described by the Fermi function. In this quasiequilibrium case the temperature of the electron system may fluctuate around its average value:\cite{heikkila09} The average temperature is determined by a heat balance, where incoming and outgoing heat flows cancel each other. These heat flows exhibit fluctuations due to the probabilistic nature of the tranmission of electrons through the tunnel barriers formed by the insulating layers. As a result, the temperature of the electron system fluctuates as well.

Recently, temperature fluctuation statistics (TFS) has been studied in non-interacting islands,\cite{heikkila09} and overheated single-electron transistors.\cite{laakso10,laakso10b} The TFS become especially interesting in SINIS structures, where a normal metallic island with a single-particle level spacing $\delta\ll\Delta$ is connected to two superconducting reservoirs via tunnel junctions.\cite{laakso12} By virtue of the interplay of regular quasiparticle tunneling and two-electron Andreev tunneling, the statistics of temperature fluctuations are strongly non-Gaussian when all transmission eigenvalues of the junctions are small, characterized by a number $\mathcal{F}\lesssim(\delta/\Delta)^{3/4}$, and when the voltage is close to $2\Delta$, $|V-2\Delta|\lesssim\sqrt{\delta\Delta}$.

\begin{figure}
	\includegraphics[width=0.7\columnwidth]{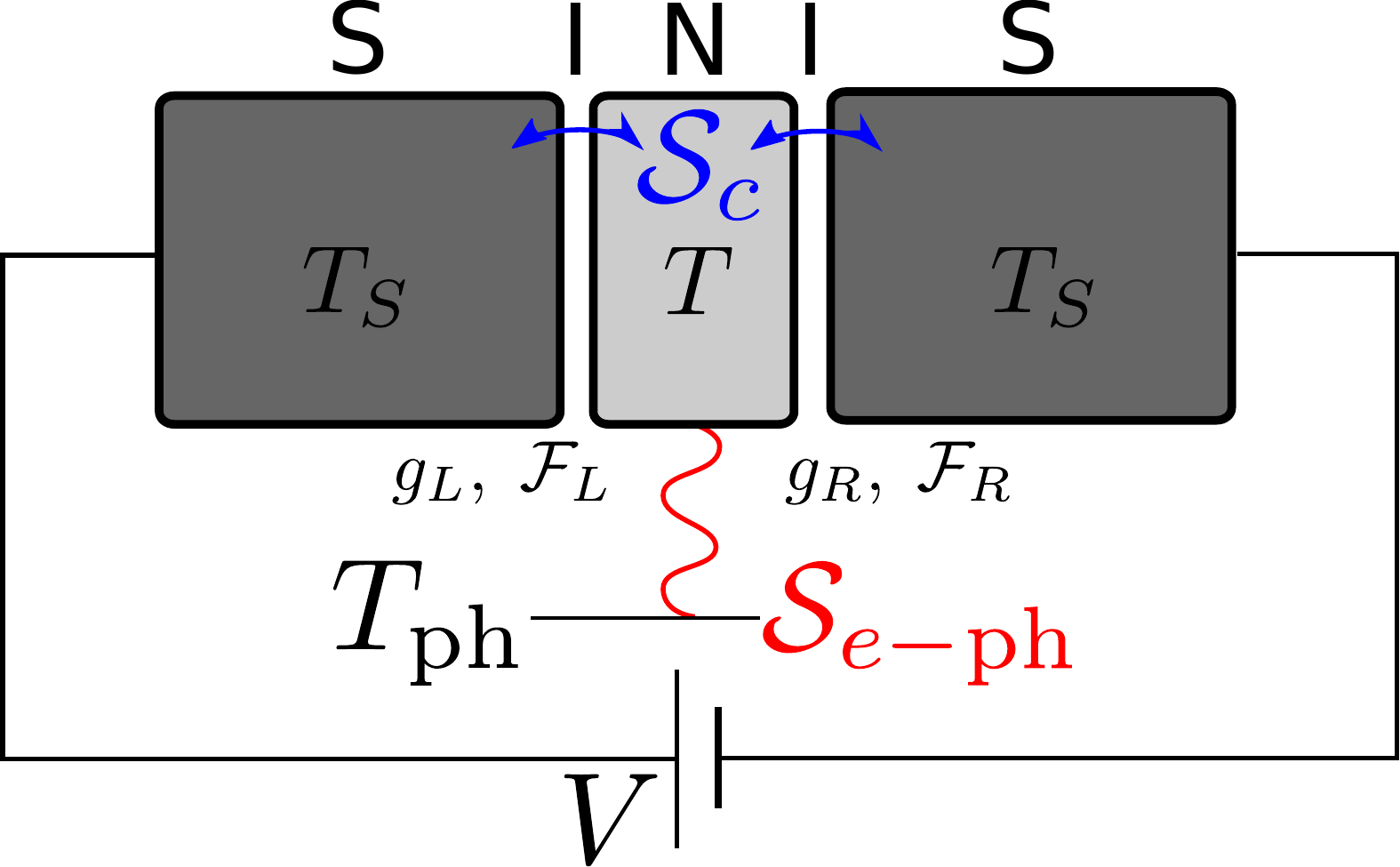}
	\caption{(color online) Schematic diagram of a SINIS structure, biased with voltage $V$. The normal metallic island is connected via tunnel contacts to superconducting reservoirs at temperature $T_S$. The transmission eigenvalue distribution of the contacts is characterized by two numbers, $g$ and $\mathcal{F}$. The temperature on the island, $T$, fluctuates due to fluctuations in the energy flows to the leads and the phonons. Tunneling through the contacts is described by the connector action $\mathcal{S}_c$, and the electron--phonon coupling with the corresponding action $\mathcal{S}_{e-\mr{ph}}$.}
	\label{fig:schema}
\end{figure}

In this Article we apply the theoretical methods developed in \refcite{laakso10b} to the SINIS structure, shown schematically in Fig.~\ref{fig:schema}. We present a detailed theoretical description of the temperature fluctuations in the Gaussian and non-Gaussian regimes and show how these fluctuations give rise to large fluctuations in the electric current, with a noise power that can exceed the intrinsic current fluctuations by a factor of a hundred. We extend the previous theoretical description to include electron--phonon coupling and show that its effect on our results is negligible, provided that the superconducting reservoirs can be kept at a relatively low temperature, $T_S\lesssim0.1\Delta$.

This Article is structured as follows: Section \ref{sec:theory} outlines the theoretical methods utilized for the calculation of the TFS. In Sec.~\ref{sec:gaussian} we focus on the temperature and current fluctuations in the Gaussian regime and in Sec.~\ref{sec:nongaussian} on the fluctuations in the non-Gaussian regime. We describe a method for the detection of these fluctuations in Sec.~\ref{sec:cf}. We conclude and discuss our findings in Sec.~\ref{sec:conclusion}. In the entirety of the text, we use units in which $\hbar=k_B=e=1$.

\section{Effective action formulation}\label{sec:theory}

We study the temperature fluctuations using the Keldysh action technique, reviewed in \refcite{kamenev09} and detailed for the application to temperature fluctuations in Refs.~\onlinecite{heikkila09} and \onlinecite{laakso10b}. We assume that the reservoirs are diffusive bulk superconductors in equilibrium, and that the normal metal island is also diffusive and in quasiequilibrium. The condition for the island to be in quasiequilibrium is given by $(G/G_Q)(G_i/G_Q)\{\delta/\max(T,V)\}\ll1$, where $G$ is the total conductance through the structure, $G_i$ is the conductance of the island not including the tunnel barriers, and $\delta$ is the single-particle level spacing on the island. The conductance quantum is defined as $G_Q=2e^2/h$. 

After incorporating the conservation of energy $E$ and charge $Q$ on the normal metal island with the aid of two Lagrange multipliers, $\xi$ and $\chi$,\cite{pilgram04} the Keldysh partition function for the SINIS structure can be written in the form
\begin{align}\label{eq:keldyshZ}
 \mathcal{Z}=&\int\mathcal{D}\xi(t)\mathcal{D}\chi(t)\mathcal{D}E(t)\mathcal{D}Q(t)\exp\left\{-\int\d t\left(\xi(t)\dot{E}(t)\right.\right. \nonumber \\ &\biggl.\left.+\chi(t)\dot{Q}(t)-\mathcal{S}(\xi,\chi,E)\right)\biggr\},
\end{align}
where $\mathcal{S}=\mathcal{S}_c+\mathcal{S}_{e-\mr{ph}}$ is the action of the system, consisting of the connector action $\mathcal{S}_c$, describing the contacts to the superconducting reservoirs, and the electron--phonon action $\mathcal{S}_{e-\mr{ph}}$, describing the coupling of electrons on the island to the lattice phonons. For the total energy and charge on the island we use a simple model where $E=\pi^2T^2/(6\delta)$ and $Q=C\mu$. Here, $T$ is the temperature, $\mu$ the chemical potential, and $C$ the capacitance of the island. The single-particle level spacing depends on the size of the island $\mathcal{V}$, and the density of states at the Fermi level $\nu$, via $\delta=1/(\nu\mathcal{V})$. For a small copper island, $\nu\approx10^{47}\:\mr{J}^{-1}\mr{m}^{-3}$, $\mathcal{V}\approx10^{-21}\:\mr{m}^3$, and aluminum leads, $\Delta\approx10^{-23}\:\mr{J}$, resulting in $\delta/\Delta\approx10^{-3}$.

An equation describing the time evolution of the probability distribution for $E$ can be obtained from Eq.~\eqref{eq:keldyshZ} in the form
\begin{equation}\label{eq:evolution}
 \frac{\partial\mathcal{P}(E,t)}{\partial t}=\hat{H}\mathcal{P}(E,t),
\end{equation}
where the operator $\hat{H}$ is obtained from $\mathcal{S}(\xi,E)$ with the substitution $\xi\mapsto-\partial/\partial E$. The operator $\hat{H}$ must be normally ordered, i.e., $\xi$ must be to the left of all $E$.\cite{laakso10b} The stationary solution of this equation yields the statistics of the fluctuations of $E$, and is therefore our main object of interest.

\subsection{Connector action}\label{subs:sc}

The connector action for a SINIS hybrid structure is very similar to that of an all-normal structure.\cite{heikkila09} The gauge transformed quasiclassical Green's functions in Keldysh ($\check{\phantom{g}}$) $\otimes$ Nambu ($\hat{\phantom{g}}$) space are given by
\begin{align}\label{eq:green}
	\check{G}=&\exp\left\{-\frac{1}{2}\left(\chi\check{\tau}_1\otimes\hat{\tau}_3+\xi\epsilon\check{\tau}_1\otimes\hat{I}-\xi\mu\check{\tau}_1\otimes\hat{\tau}_3\right)\right\}\check{G}_0 \\ \nonumber &\times\exp\left\{\frac{1}{2}\left(\chi\check{\tau}_1\otimes\hat{\tau}_3+\xi\epsilon\check{\tau}_1\otimes\hat{I}-\xi\mu\check{\tau}_1\otimes\hat{\tau}_3\right)\right\}, \\
	\check{G}_0=&\begin{pmatrix}
			\hat{G}^R & \hat{G}^K \\
			0 & \hat{G}^A \\
	\end{pmatrix}, \\
	\hat{G}^R=&\frac{\sgn(\epsilon)}{\sqrt{\epsilon^2-\Delta^2}}\left(\epsilon\hat{\tau}_3+\ii\hat{\tau}_2\Delta\right), \\
	\hat{G}^A=&-\hat{\tau}_3\hat{G}^{R\dagger}\hat{\tau}_3, \\
	\hat{G}^K=&\hat{G}^R\hat{h}-\hat{h}\hat{G}^A,
\end{align}
where $\tau_i$ are Pauli matrices, $\hat{h}=\mr{diag}(1-2f(\epsilon),2f(-\epsilon)-1)$, and $f(\epsilon)$ is the Fermi function.

In terms of these functions the connector action is given by\cite{snyman08}
\begin{equation}
 \mathcal{S}_c=\frac{1}{2}\sum_\alpha\sum_{n\in\alpha}\int\frac{\d\epsilon}{2\pi}\Tr\ln\left[1+T_n^\alpha\frac{\{\check{G}_\alpha,\check{G}_I\}-2}{4}\right],
\end{equation}
where $\Tr$ denotes a trace over Keldysh and Nambu indices, and $T_n^\alpha$ are the transmission eigenvalues of contact $\alpha\in L,R$. When some parts of the structure are superconducting, this quantity becomes relatively complicated due to the matrix structure in Nambu space. In the tunneling limit $T_n^\alpha\ll1$, however, this expression can be considerably simplified. Expanded to the second order in the transmission eigenvalues it reads
\begin{align}\label{eq:sc}
 \mathcal{S}_c=&\frac{1}{8}\sum_\alpha g_{\alpha}\int\frac{\d\epsilon}{2\pi}\Tr\left[\{\check{G}_\alpha,\check{G}_I\}-2\right] \\ \nonumber &-\frac{1}{64}\sum_\alpha g_{\alpha}\mathcal{F}_\alpha\int\frac{\d\epsilon}{2\pi}\Tr\left[\left(\{\check{G}_\alpha,\check{G}_I\}-2\right)^2\right],
\end{align}
where $g_\alpha=\sum_n T_n^\alpha\ll1$ is the dimensionless conductance of contact $\alpha$ and $\mathcal{F}_\alpha=\sum_n(T_n^\alpha)^2/\sum_n T_n^\alpha\ll1$ a parameter describing the distribution of transmission eigenvalues. Typical value is $\mathcal{F}_\alpha\simeq 10^{-5}$--$10^{-6}$ for aluminum junctions.\cite{greibe11} The expansion to the second order is necessary since the first order term is exponentially suppressed at bias voltages below $V=2\Delta$. The first term describes tunneling of quasiparticles through the insulating barrier, whereas the second term describes tunneling of Cooper pairs (Andreev tunneling) and processes where two quasiparticles are transmitted simultaneously through the barrier.

In the incoherent limit, neglecting the proximity effect on the island, the connector action for the left and right contacts can be evaluated independently, in such a way that the superconductor always has zero chemical potential and the island has $\mu=-\mu_{L,R}$ for the left and right contact, respectively, and $V=\mu_L-\mu_R$. This circumvents the need to take into account the rotating phase of the superconducting order parameter in superconductors at a finite voltage.

We proceed by calculating the connector action in two parts. Carrying out the trace in the first term in Eq.~\eqref{eq:sc} results in
\begin{align}\label{eq:sc1}
 \mathcal{S}_c^{(1)}=&\sum_\alpha g_{\alpha}\int\frac{\d\epsilon}{2\pi}\mr{Re}\left[g(\epsilon)\right] \nonumber \\ &\times\left[2(e^{\chi+\xi(\epsilon+\mu_\alpha)}-1)f_\alpha(\epsilon)(1-f_I(\epsilon+\mu_\alpha))\right. \nonumber \\ &\left.+2(e^{-\chi-\xi(\epsilon+\mu_\alpha)}-1)f_I(\epsilon+\mu_\alpha)(1-f_\alpha(\epsilon))\right].
\end{align}
Here, $\mr{Re}\left[g(\epsilon)\right]=\mr{Re}\left[|\epsilon|/\sqrt{\epsilon^2-\Delta^2}\right]$, is the BCS density of states. The second term in Eq.~\eqref{eq:sc} contains four kinds of terms: Terms proportional to $\mr{Re}\left[g(\epsilon)\right]$, $(\mr{Re}\left[g(\epsilon)\right])^2$, $(\mr{Re}\left[F(\epsilon)\right])^2$, and $|F(\epsilon)|^2$, where $F(\epsilon)=\Delta\sgn(\epsilon)/\sqrt{\epsilon^2-\Delta^2}$. The first three are non-zero only at energies $|\epsilon|\geq\Delta$, and can be neglected in comparison to Eq.~\eqref{eq:sc1} due to the small prefactor $\mathcal{F}_\alpha$, whereas terms of the last kind are non-zero everywhere and therefore important. Disregarding the other terms brings the second term in Eq.~\eqref{eq:sc} into the form
\begin{align}\label{eq:sc2}
 \mathcal{S}_c^{(2)}=&\frac{1}{4}\sum_\alpha g_{\alpha}\mathcal{F}_\alpha\int\frac{\d\epsilon}{2\pi}|F(\epsilon)|^2 \nonumber \\ &\times\left[(e^{2\chi+2\xi\mu_\alpha}-1)(1-f_I(\epsilon+\mu_\alpha))(1-f_I(-\epsilon+\mu_\alpha))\right. \nonumber \\ &\left.+(e^{-2\chi-2\xi\mu_\alpha}-1)f_I(\epsilon+\mu_\alpha)f_I(-\epsilon+\mu_\alpha)\right].
\end{align}
Using the connector action we can evaluate the charge and heat current, $I=\partial_\chi \mathcal{S}_c|_{\chi=0}$, $\dot{H}=\partial_\xi \mathcal{S}_c|_{\xi=0}$, and their noise powers $S_I=\partial^2_\chi \mathcal{S}_c|_{\chi=0}$, $S_{\dot{H}}=\partial^2_\xi \mathcal{S}_c|_{\xi=0}$.

\subsection{Heat current near $V=2\Delta$}\label{subs:hc}

\begin{figure}
	\includegraphics[width=0.9\columnwidth]{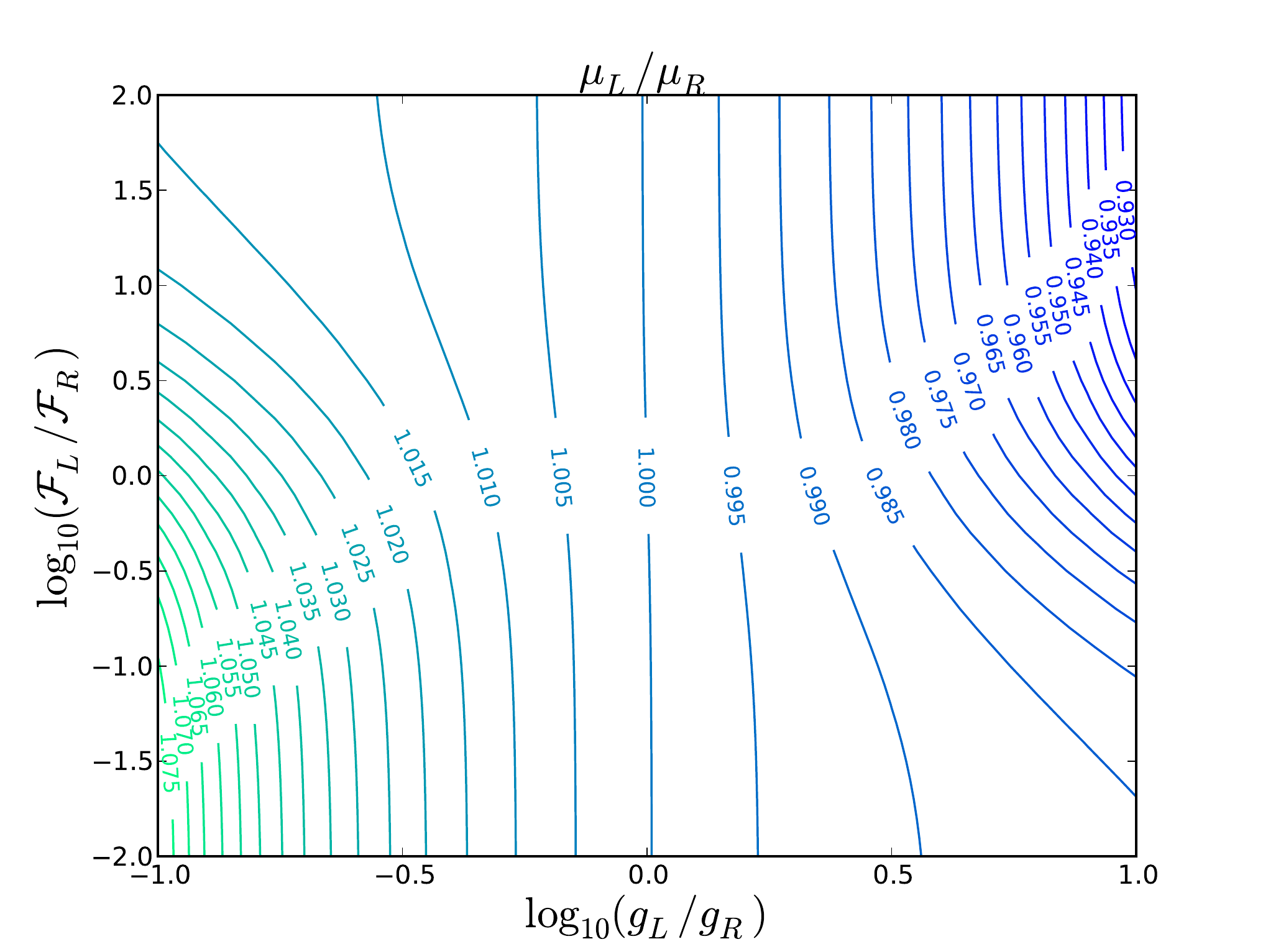}
	\caption{(color online) Ratio of the voltage drops across left and right junctions as a function of conductance and $\mathcal{F}$ asymmetries for $V=2\Delta$. Note the logarithmic scales.}
	\label{fig:asymmetry}
\end{figure}

For simplicity, we consider a symmetric structure with $g_L=g_R=g$ and $\mathcal{F}_L=\mathcal{F}_R=\mathcal{F}$, and therefore $\mu_L=-\mu_R=V/2$. This is not a restrictive assumption: Moderate asymmetries in the properties of the junctions does not affect our results near $V=2\Delta$. The reason for this is the strong nonlinearity of the current-voltage relation in the vicinity of $V=2\Delta$, which guarantees a small asymmetry in the voltage drops across the junctions (see Fig.~\ref{fig:asymmetry}). We also assume that the electric $RC$-time of the structure is small compared to the characteristic time scale for energy transport. In this case the evolution of $\chi(t)$ and $\mu(t)$ are well-described by their saddle point solutions,\cite{heikkila09} and when the structure is symmetric, $\chi(t)=0$. Furthermore, we can neglect cumulants of the third order and higher in the quasiparticle heat current, since they are small at low temperatures. For this reason we expand the first part of the connector action to the second order in $\xi$. The total connector action becomes $\mathcal{S}_c=\sum_\alpha(\dot{H}_\alpha^{(1)}\xi+S_{\dot{H},\alpha}^{(1)}\xi^2/2)+\mathcal{S}_c^{(2)}$, where $\dot{H}_\alpha^{(1)}$ and $S_{\dot{H},\alpha}^{(1)}$ are the energy current due to quasiparticle tunneling through contact $\alpha$ and its noise. For zero temperature leads, terms in $\mathcal{S}_c^{(1)}$ can be evaluated analytically with the aid of formulas from \refcite{anghel01}. For example, when $(\Delta+V/2)/T\gg(\Delta-V/2)/T$,
\begin{align}
 \dot{H}^{(1)}=&-\frac{2g}{\pi}\frac{T^{5/2}}{\sqrt{2\Delta}}\left\{\frac{\Delta}{T}\left[\Gamma\left(\frac{3}{2}\right)g_{3/2}(a)+\Gamma\left(\frac{1}{2}\right)ag_{1/2}(a)\right]\right. \nonumber \\ &+\left.\frac{3}{4}\left[\Gamma\left(\frac{5}{2}\right)g_{5/2}(a)+\Gamma\left(\frac{3}{2}\right)ag_{3/2}(a)\right]\right\}, \\ S_{\dot{H}}^{(1)}=&\frac{2g}{\pi}\frac{T^{7/2}}{\sqrt{2\Delta}}\left\{\frac{\Delta}{T}\left[\Gamma\left(\frac{5}{2}\right)g_{5/2}(a)+2\Gamma\left(\frac{3}{2}\right)ag_{3/2}(a)\right.\right.\nonumber \\ &\left.+\Gamma\left(\frac{1}{2}\right)a^2g_{1/2}(a)\right]+\frac{3}{4}\left[\Gamma\left(\frac{7}{2}\right)g_{7/2}(a)\right.\nonumber \\ &\left.\left.+2\Gamma\left(\frac{5}{2}\right)ag_{5/2}(a)+\Gamma\left(\frac{3}{2}\right)a^2g_{3/2}(a)\right]\right\},
\end{align}
where $g_l(a)=-\mr{Li}_l(-e^{-a})$, $\mr{Li}_l(x)$ being the $l$th order polylogarithm, and $a=(\Delta-V/2)/T$. For leads at finite temperature $T_S$, the heat current includes an additional term 
\begin{equation}
 +\frac{2g}{\pi}\sqrt{\frac{2\pi T_S}{\Delta}}\exp(-\Delta/T_S)\Delta^2.
\end{equation}
Due to the exponential dependence on $T_S$ we neglect this term in the following, and thus our zero-temperature approximation is valid as long as $T_S\ll\Delta$.

At zero temperature $\mathcal{S}_c^{(2)}$ can be evaluated to
\begin{align}\label{eq:andreev}
 \mathcal{S}_c^{(2)}=&g\mathcal{F}\frac{\Delta}{2\pi}\mr{arctanh}(V/2\Delta)\left(e^{\xi V}-1\right), \nonumber \\ &\approx g\mathcal{F}\frac{\Delta}{4\pi}\ln\left[\frac{2}{1-V/(2\Delta)}\right]\left(e^{\xi V}-1\right).
\end{align}
Finite temperature gives rise to logarithmic corrections which can be disregarded in the following. As seen from the second line, the action has a logarithmic divergence at the threshold $V=2\Delta$, an artifact of the expansion of the action in transmission eigenvalues. A careful analysis shows that this divergence is cut off in such a way that the argument of the logarithm goes to $\mathcal{F}^{-1}$ at $V=2\Delta$. However, the exact functional form of the action near the threshold is not necessary. Instead, the strength of the Andreev tunneling can be described by a single parameter $\Gamma_A$ in the form $\mathcal{S}_c^{(2)}=\Gamma_A\left(e^{\xi V}-1\right)$.

\subsection{Electron--phonon action}\label{subs:seph}

Elecron--phonon coupling in the lowest order is described with the action (see Appendix \ref{app:derivation})
\begin{align}\label{eq:e-ph}
 \mathcal{S}_{e-\mr{ph}}=&\frac{\Sigma\mathcal{V}}{24\zeta(5)}\int_0^\infty \d\epsilon \epsilon^3\left[\left(e^{\xi\epsilon}-1\right)n(\epsilon,T_\mr{ph})(1+n(\epsilon,T))\right. \nonumber \\ &\left.+\left(e^{-\xi\epsilon}-1\right)n(\epsilon,T)(1+n(\epsilon,T_\mr{ph}))\right],
\end{align}
where $n(\epsilon,T)$ is the Bose distribution function, $\Sigma$ the electron--phonon coupling constant, $\mathcal{V}$ the volume of the island, and $\zeta(5)\approx1.04$. The electron--phonon coupling strength can be adjusted by a dimensionless parameter $g_{e-\mr{ph}}=\Sigma\mathcal{V}\Delta^3$, which can be compared to the dimensionless conductance of the junctions. To cast this into a more understandable form, we write
\begin{align*}
	g_{e-\mr{ph}}/g=&\frac{\Sigma}{2\cdot10^9\:\mr{WK}^{-5}\mr{m}^{-3}}\frac{\mathcal{V}}{2.8\cdot10^{-21}\:\mr{m}^3}\left(\frac{\Delta/k_B}{2.0\:\mr{K}}\right)^3 \\
	&\times\left(\frac{R_T}{12.9\:\mr{k\Omega}}\right).
\end{align*}
For example, for a small copper island and $G=G_Q$, $g_{e-\mr{ph}}/g=1$. When $g_{e-\mr{ph}}/g\ll1$ we can disregard the phonons altogether.

For our purposes it is again enough to calculate Eq.~\eqref{eq:e-ph} to the second order in $\xi$. For phonons at zero temperature this can be done analytically. The electron--phonon heat current is given by
\begin{equation}
	\dot{H}_{e-\mr{ph}}=-\frac{g_{e-\mr{ph}}}{\Delta^3}T^5,
\end{equation}
and the heat current noise by
\begin{equation}
	S_{\dot{H},e-\mr{ph}}=\frac{\pi^6g_{e-\mr{ph}}}{189\zeta(5)\Delta^3}T^6.
\end{equation}
Finite temperature phonons add a term 
\begin{equation}
 +\frac{g_{e-\mr{ph}}}{\Delta^3}T_\mr{ph}^5
\end{equation} 
to the heat current. Due to the high powers of $T$ these are negligibly small at low temperatures for not extremely large $g_{e-\mr{ph}}/g$.

\section{Gaussian regime}\label{sec:gaussian}

The temperature fluctuations are Gaussian when $\delta\ll\mathcal{F}^{4/3}\Delta$ or $|V-2\Delta|\gg\sqrt{\delta\Delta}$. In this case the temperature distribution is characterized simply by its average value and variance, which can be solved from the heat current and its noise in a straightforward manner.

\subsection{Average temperature}

\begin{figure}
	\includegraphics[width=\columnwidth]{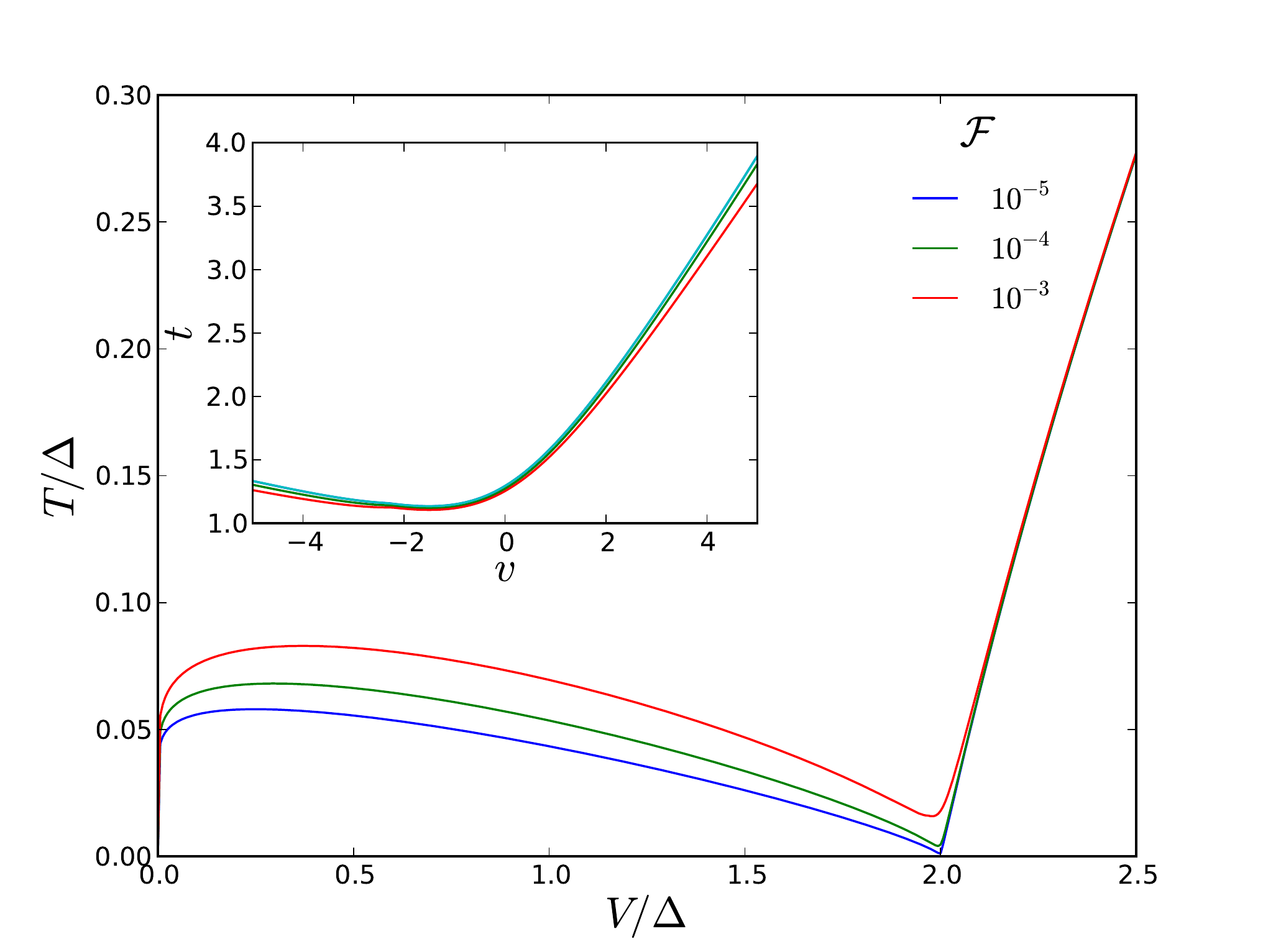}
	\caption{(color online) Average temperature of the normal metallic island in a SINIS setup for some values of $\mathcal{F}$. The inset shows that the scaled temperature $t$ is independent of $\mathcal{F}$ near the scaled voltage $v=0$.}
	\label{fig:avgt}
\end{figure}

The average temperature as a function of bias voltage, assuming zero-temperature leads and a vanishing electron--phonon coupling, is shown in Fig.~\ref{fig:avgt}. At low bias voltages the SINIS system exhibits overheating due to the second order Andreev processes.\cite{rajauria08} The sharp rise in temperature can be described analytically: Solving for $T$ from $\dot{H}=0$ in the limit $(\Delta-V/2)/T\gg1$, $V\to0$ yields the equilibrium temperature
\begin{equation}
	T\simeq\frac{\Delta}{\ln\left[(\Delta/V)^2/\mathcal{F}\right]}.
\end{equation}
As the voltage is increased, the cooling effect gets stronger, and the temperature reaches its minimum near $V=2\Delta$. We can again solve for the equilibrium temperature, which at $V=2\Delta$ is given by
\begin{equation}\label{eq:eqt}
	T=\left(\frac{\tilde{\gamma}}{\Gamma\left(\frac{3}{2}\right)g_{3/2}(0)}\right)^{2/3}\Delta,
\end{equation}
where $\tilde{\gamma}=\sqrt{2}\pi\Gamma_A/(g\Delta)$ and $g_{3/2}(0)\approx0.77$. The minimum temperature is set by the magnitude of the Andreev processes. Above $V=2\Delta$ the temperature is set mainly by the first order tunneling and is essentially independent of $\Gamma_A$: $T\propto V/2-\Delta$.

We point out that instead of including the second order contributions the heating effect can also be reproduced by a finite broadening of the BCS density of states with the replacement $\epsilon\mapsto\epsilon+\ii\eta$ in the superconductor Green's function.\cite{pekola04} The physical origin of this imaginary term can be, for example, the electromagnetic environment of the tunnel structure.\cite{pekola10}

\subsection{Gaussian fluctuations}

\begin{figure}
	\includegraphics[width=\columnwidth]{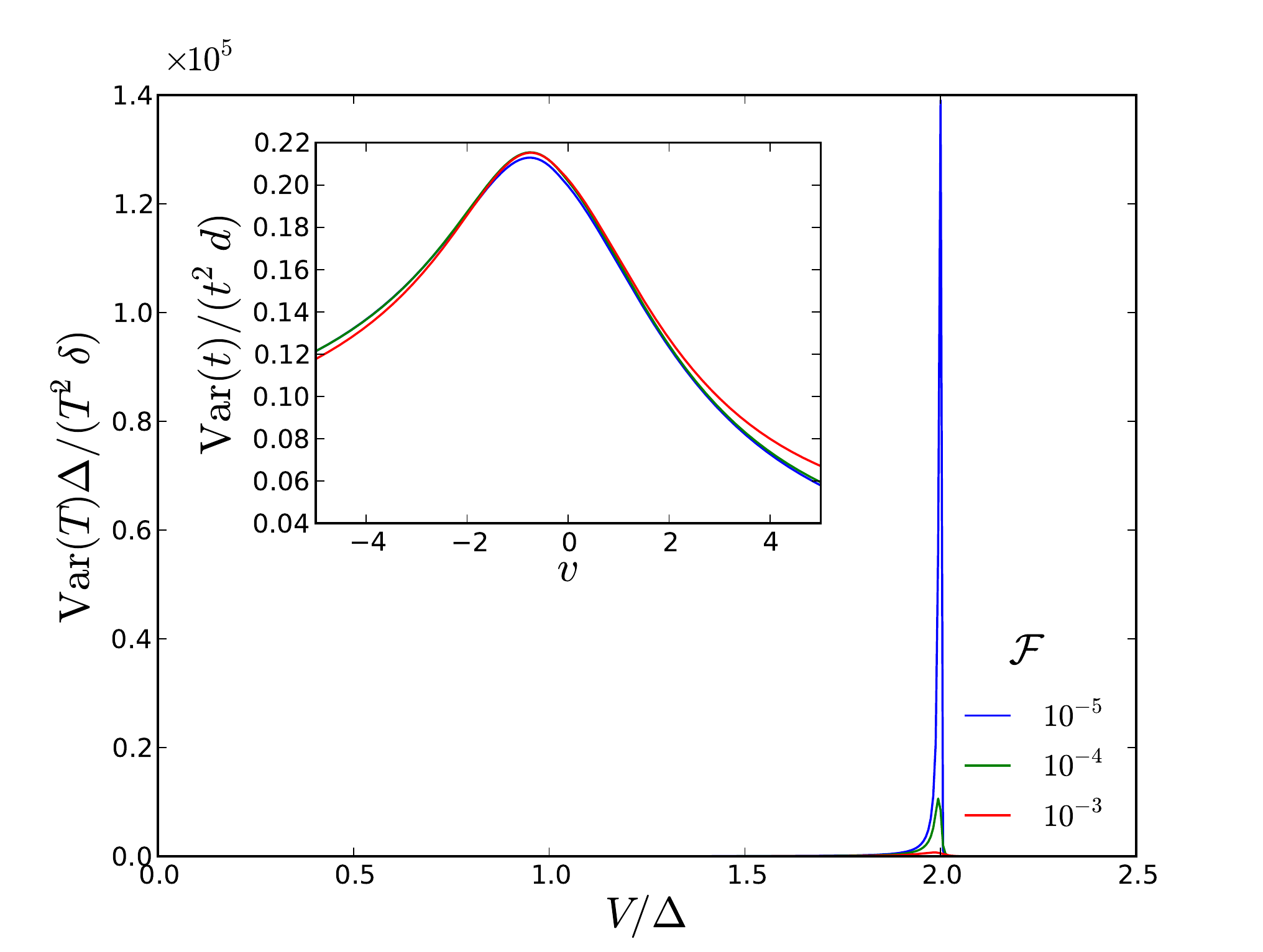}
	\caption{(color online) Variance of the temperature of the normal metallic island in a SINIS setup for some values of $\mathcal{F}$. The inset shows that the variance of the scaled temperature $t$ is independent of $\mathcal{F}$ near the scaled voltage $v=0$.}
	\label{fig:vart}
\end{figure}

The variance of the temperature, assuming a Gaussian distribution, can be estimated by \cite{heikkila09}
\begin{equation}
 \mathrm{Var}(T)=\frac{S_{\dot{H}}}{2G_\mr{th}C},
\end{equation}
where $C=\pi^2T/(3\delta)$ is the heat capacity of the island and $G_\mr{th}$ is the heat conductance. The normalized variance as a function of bias voltage in the Gaussian approximation is shown in Fig.~\ref{fig:vart}. It is strongly peaked around the threshold voltage $V\approx2\Delta$, and falls rapidly when the voltage is tuned away from this threshold.

Motivated by the $\tilde{\gamma}^{2/3}$ dependence of the minimum temperature we define scaled variables $t=T/(\Delta\tilde{\gamma}^{2/3})$, $v=(V/2-\Delta)/(\Delta\tilde{\gamma}^{2/3})$, and $d=\delta/(\Delta\tilde{\gamma}^{4/3})$, effectively eliminating $\tilde{\gamma}$ from the formulas in the limit $\tilde{\gamma}\ll1$. In terms of these scaled variables, we obtain
\begin{equation}\label{eq:vart}
 \mathrm{Var}(t)=\frac{3d}{\pi^{5/2}t^{3/2}\left(\frac{3}{4}g_{3/2}(a)+ag_{1/2}(a)+a^2g_{-1/2}(a)\right)},
\end{equation}
where $a=-v/t$. For $V=2\Delta$, i.e., $a=0$, we get using Eq.~\eqref{eq:eqt},
\begin{equation}
 \mathrm{Var}(t)=\frac{2d}{\pi^2}.
\end{equation}
Since the temperature is limited from below, the distribution cannot be completely Gaussian. The low-temperature tail is always non-Gaussian. When the standard deviation of the distribution approaches the expectation value, also the main body of the distribution becomes non-Gaussian. Therefore, for $d\gtrsim\langle t\rangle\approx1$, temperature fluctuations around $V=2\Delta$ become non-Gaussian and of the same order of magnitude as the average temperature, rendering Eq.~\eqref{eq:vart} incorrect.

\subsection{Induced current fluctuations}

\begin{figure}
	\includegraphics[width=\columnwidth]{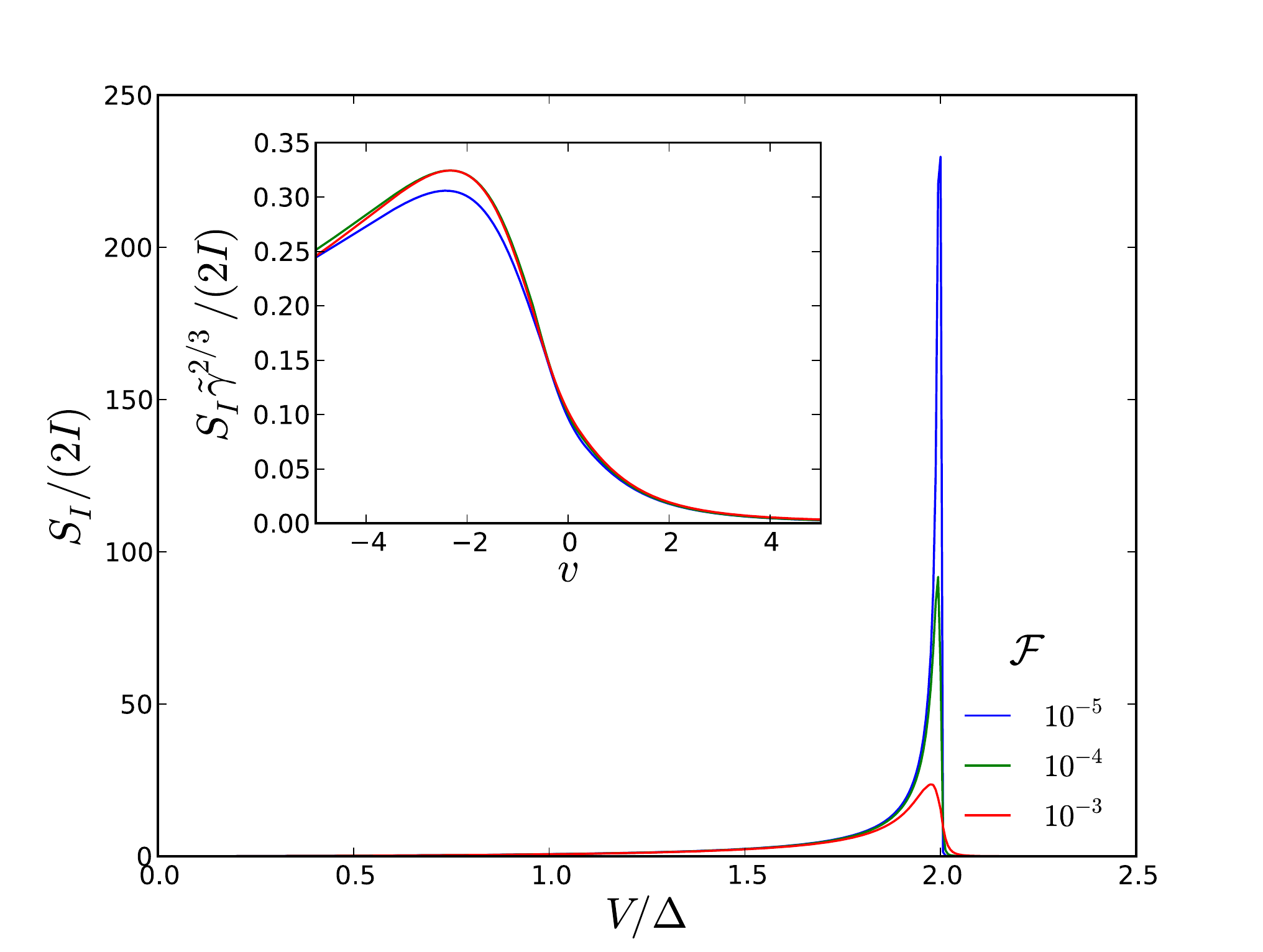}
	\caption{(color online) Fano factor of the temperature fluctuation induced current noise for some values of $\mathcal{F}$. Inset shows the $\tilde{\gamma}^{-2/3}$ scaling of the Fano factor near the scaled voltage $v=0$.}
	\label{fig:tfluctfano}
\end{figure}

Temperature fluctuations typically cause low-frequency fluctuations in the electric current.\cite{laakso10,laakso10b} The current noise due to temperature fluctuations is given by\cite{laakso10}
\begin{equation}
 S_{I,\mr{ind.}}=\left(\frac{\partial I/\partial T}{\partial \dot{H}/\partial T}\right)^2 S_{\dot{H}},
\end{equation}
and is best characterized with the Fano factor of the SINIS system, $F_\mr{SINIS}=S_I/(2I)$. Near $V=2\Delta$ we can use the analytical formulas for these quantities to obtain $S_{I,\mr{ind.}}/I\approx c\tilde{\gamma}^{-2/3}\propto t^{-1}$, where the numerical factor $c\approx0.17$. This strongly peaked character of the Fano factor is shown in Fig.~\ref{fig:tfluctfano}. For low $\mathcal{F}$, the Fano factor can reach values of several hundreds, two orders of magnitude larger than the Fano factor $F_\mr{SINIS}\approx1$ due to intrinsic current fluctuations in the NIS junctions.

\subsection{Effect of non-zero temperature and electron--phonon coupling}

\begin{figure}
	\includegraphics[width=\columnwidth]{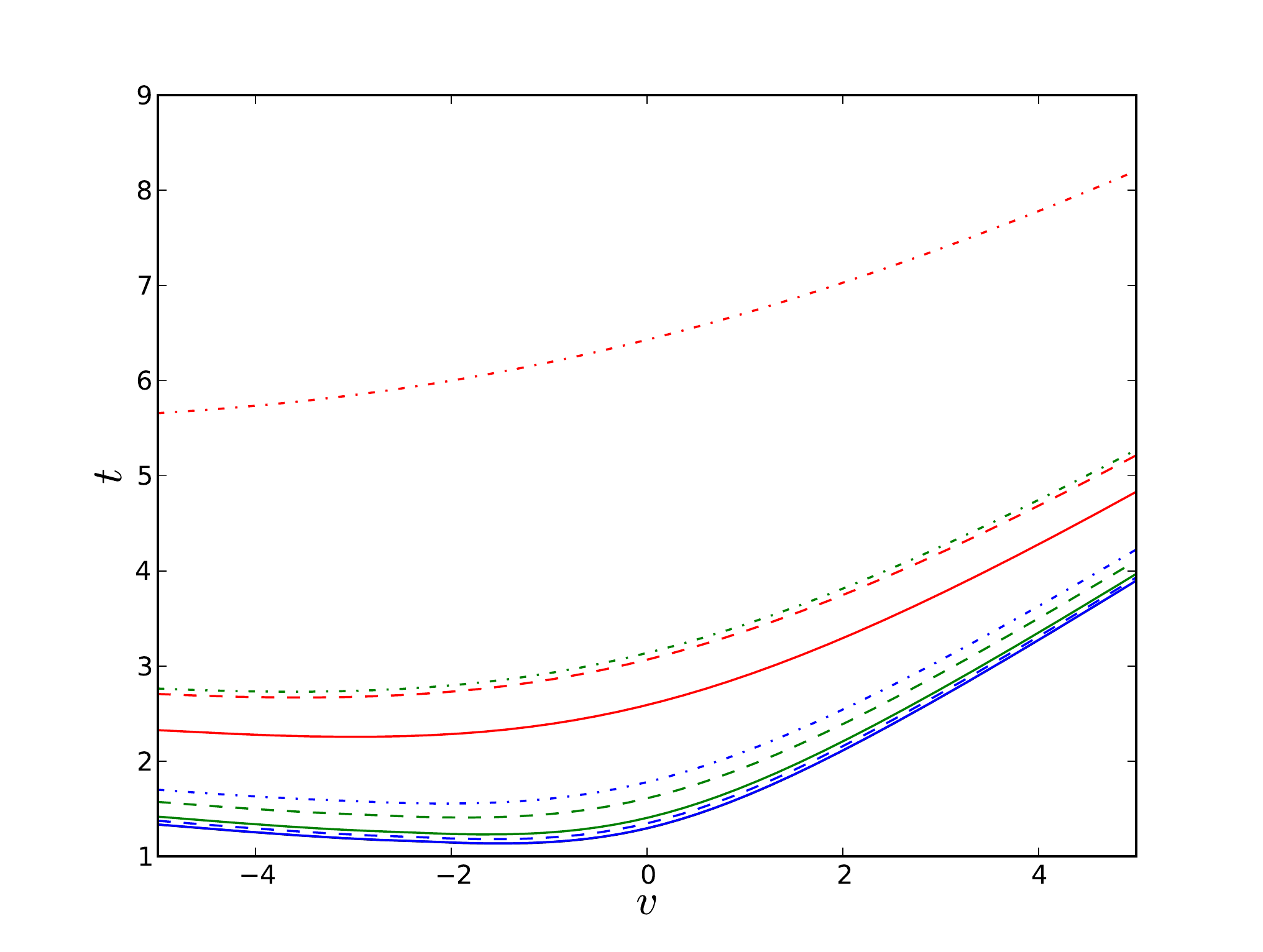}
	\caption{(color online) Average temperature of the normal metallic island in a SINIS setup for $\mathcal{F}=10^{-5}$, different values of $T_S=T_\mr{ph}$: $0.06\Delta$ (blue), $0.08\Delta$ (green), and $0.1\Delta$ (red), and different values of $g_{e-\mr{ph}}/g$: $0$ (solid lines), $1$ (dashed lines), and $10$ (dash-dotted lines). The reference line for $T_S=T_\mr{ph}=0$, $g_{e-\mr{ph}}/g=0$ is shown in black, which is lost under the solid blue line.}
	\label{fig:avgt-ts-gph}
\end{figure}

Non-zero temperature in the leads and/or an appreciable electron--phonon coupling have an effect on the expectation value and variance of the temperature on the island. The effect is strongest in the overheating regime, where the temperature of the island is highest. We are mostly interested in the effect around $V=2\Delta$, however, since that is where the non-Gaussian effects should appear. The effect to the expectation value of temperature in the scaled units around $v=0$ is shown in Fig.~\ref{fig:avgt-ts-gph}. Once the temperature of the leads approaches $0.1\Delta$, the cooling effect begins to diminish. Similarly, once the phonon temperature approaches $0.1\Delta$ and the electron--phonon coupling parameter approaches unity, the cooling effect begins to diminish. The effect on the scaled variance is easily inferred from the effect on the expectation value: Since the variance is proportional to $t^{-3/2}$, an increase of temperature by a factor of $2$ around $v=0$ decreases the maximum variance by a factor of $0.35$.

\section{Non-Gaussian statistics}\label{sec:nongaussian}

\subsection{Fokker--Planck equation}

To go beyond the Gaussian approximation we must turn our attention to the solution of Eq.~\eqref{eq:evolution}. Using the equations derived in Subs.~\ref{subs:hc} and \ref{subs:seph} the stationary state Fokker--Planck equation can be written in the form
\begin{align}
 0=&-\partial_E\left\{\dot{H}^{(1)}\mathcal{P}(E)-\partial_E\left[\frac{1}{2}S_{\dot{H}}^{(1)}\mathcal{P}(E)\right]\right\}\nonumber \\ &+\Gamma_A\left(e^{-V\partial_E}-1\right)\mathcal{P}(E) \nonumber \\ &-\partial_E\left\{\dot{H}_{e-\mr{ph}}\mathcal{P}(E)-\partial_E\left[\frac{1}{2}S_{\dot{H},e-\mr{ph}}\mathcal{P}(E)\right]\right\}.
\end{align}
The terms on the first line originate from the tunneling of individual quasiparticles, second line describes two-electron Andreev processes, and the third line electron--phonon interaction. When the average energy on the island is large compared to voltage, we can expand $e^{-V\partial_E}$ to the second order in the Andreev term, and get
\begin{equation}\label{eq:fp}
 0=-\partial_E\left\{\dot{H}\mathcal{P}(E)-\partial_E\left[\frac{1}{2}S_{\dot{H}}\mathcal{P}(E)\right]\right\},
\end{equation}
where $\dot{H}=\dot{H}^{(1)}+\dot{H}^{(2)}+\dot{H}_{e-\mr{ph}}$ and $S_{\dot{H}}=S_{\dot{H}}^{(1)}+S_{\dot{H}}^{(2)}+S_{\dot{H},e-\mr{ph}}$. Using $E=\pi^2T^2/(6\delta)$ we can write a corresponding equation for the probability distribution of temperature. This has the solution \cite{laakso10b}
\begin{equation}
 	\mathcal{P}_\mr{st}(T)=C\exp\left(\int\d T\frac{\frac{2}{3}\pi^2T\dot{H}(T)-\delta T\partial_{T}\left(\frac{S_{\dot{H}}(T)}{T}\right)}{\delta S_{\dot{H}}(T)}\right),
\end{equation}
where $C$ is a normalization constant. In the limit $\Gamma_A\ll\Delta$ we can neglect $S_{\dot{H}}^{(1)}$, $\dot{H}_{e-\mr{ph}}$, and $S_{\dot{H},e-\mr{ph}}$. These terms come with prefactors $\tilde{\gamma}^{5/3}$, $\tilde{\gamma}^{10/3}$, and $\tilde{\gamma}^{12/3}$, respectively, whereas the other three terms have a common prefactor of $\tilde{\gamma}$. In the dimensionless variables introduced in the previous section the distribution simplifies to
\begin{align}
  \mathcal{P}_\mr{st}(t)=&C\exp\left\{\int\d t\left[-\frac{\pi^{5/2}}{3\sqrt{2}d}t^{5/2}\left(\frac{1}{2}g_{3/2}(a)+ag_{1/2}(a)\right)\right.\right. \nonumber \\ &\left.\left.+\frac{\pi^2t}{3d}+\frac{1}{t}\right]\right\}.
\end{align}
For $a=0$ this becomes
\begin{equation}
  \mathcal{P}_\mr{st}(t)=Ct\exp\left\{-\frac{\pi^{5/2}}{21\sqrt{2}d}t^{7/2}g_{3/2}(0)+\frac{\pi^2t^2}{6d}\right\},
\end{equation}
which describes a distribution with an essentially Gaussian body and non-Gaussian tails. The condition of large average energy on the island translates to $d\ll1$. For $d\gtrsim1$ Eq.~\eqref{eq:fp} is no longer valid and we must take the higher derivatives into account in the Andreev term.

Continuing in the limit $\Gamma_A\ll\Delta$, and thus neglecting $S_{\dot{H}}^{(1)}$, $\dot{H}_{e-\mr{ph}}$, and $S_{\dot{H},e-\mr{ph}}$, we write
\begin{align}\label{eq:fpfinal}
 0=&-\partial_E\left[\dot{H}^{(1)}\mathcal{P}(E)\right]+\Gamma_A\left(e^{-V\partial_E}-1\right)\mathcal{P}(E), \nonumber \\ =&-\partial_E\left[\dot{H}^{(1)}\mathcal{P}(E)\right]+\Gamma_A\left[\mathcal{P}(E-V)-\mathcal{P}(E)\right].
\end{align}
Its solution depends on $V$ and $\Gamma_A$, resulting in six qualitatively different regimes that are studied in detail in \refcite{laakso12}. In general, for $\Gamma_A\lesssim g\Delta^{1/4}\delta^{3/4}$, or, equivalently, $\mathcal{F}\lesssim(\delta/\Delta)^{3/4}$, and $|V-2\Delta|\lesssim\sqrt{\delta\Delta}$, the resulting probability distribution is non-Gaussian.

\subsection{Langevin equation}

The Fokker--Planck equation, Eq.~\eqref{eq:fpfinal}, can be equivalently presented in the form of a Langevin equation. Starting from the partition function, Eq.~\eqref{eq:keldyshZ}, and using $\mathcal{S}(\xi,E)=\dot{H}^{(1)}\xi+\Gamma_A(e^{\xi V}-1)$ as the action, we can expand the exponentiated action in powers of $\Gamma_A$:
\begin{align}
 \mathcal{Z}=&\int\mathcal{D}\xi\mathcal{D}E\sum_{m=0}^\infty\frac{\Gamma_A^m}{m!}\prod_{k=1}^m\int\d t_ke^{-\int\d t\Gamma_A}\nonumber \\ &\times\exp\left\{-\int\d t\xi\left[\dot{E}-\dot{H}^{(1)}-V\sum_k\delta(t-t_k)\right]\right\}.
\end{align}
By performing now the functional integral over $\xi$ we end up with a delta functional of the quantity in square brackets. This means that only paths which satisfy the Langevin equation
\begin{equation}
 \partial_t{E}=\dot{H}^{(1)}+V\sum_k\delta(t-t_k),
\end{equation}
contribute to the partition function. The energy is increased in increments of $V$ due to Andreev events at random times $t_k$, and between the Andreev events energy changes deterministically as dictated by the energy flow $\dot{H}^{(1)}$ due to single particle processes. The Andreev events are Poisson distributed as can be seen from the statistical weighing factor in the functional integral.

Using the Langevin equation we can simulate timelines of energy, which are sometimes more informative than the full distributions obtained from the Fokker--Planck equation. These timelines are presented in \refcite{laakso12}.

\subsection{Non-Gaussian current fluctuations}

\begin{figure}
	\includegraphics[width=\columnwidth]{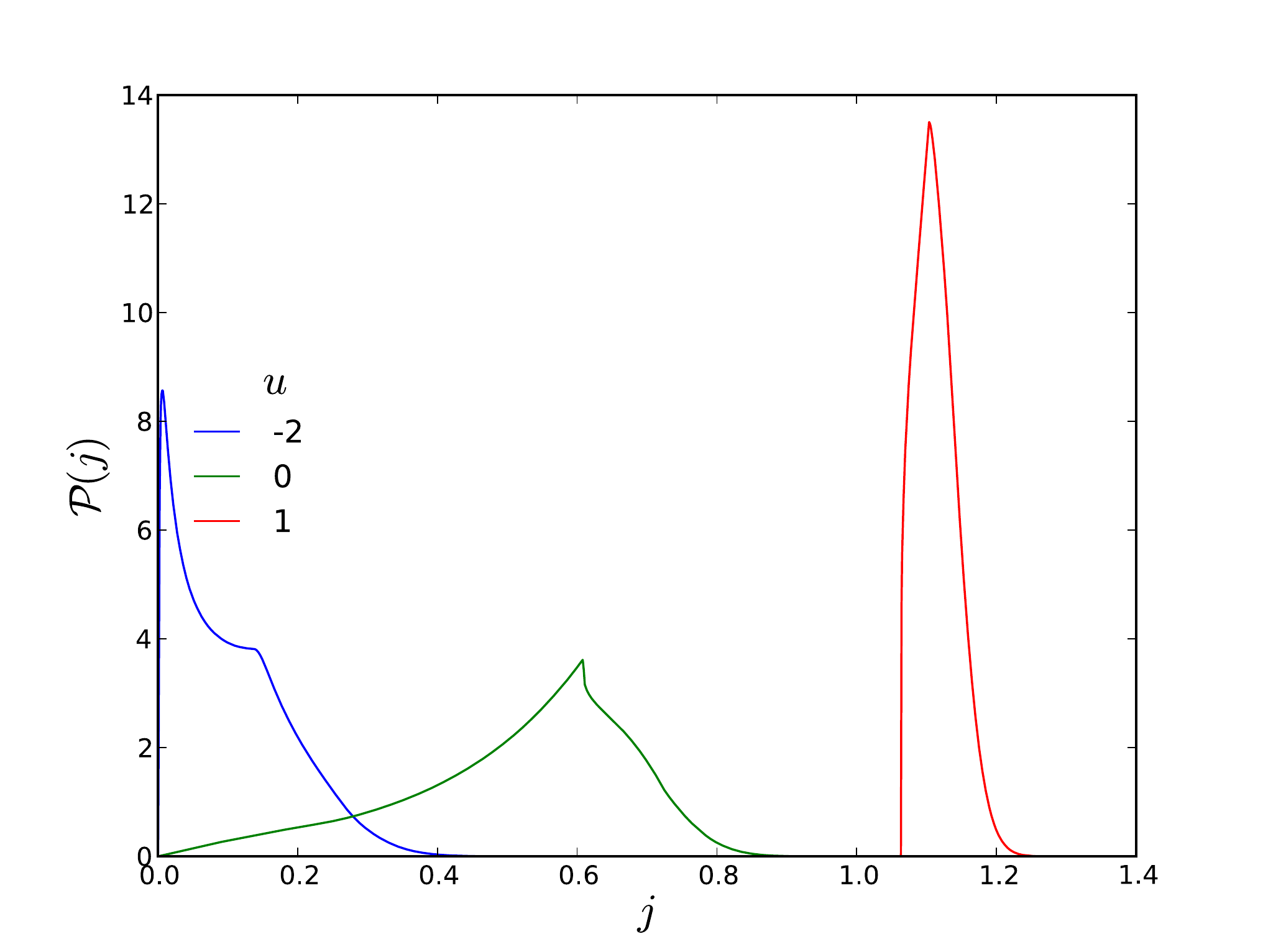}
	\caption{(color online) Probability distribution for the dimensionless current $j=\tau_rT_tI/\Delta$ in the non-Gaussian regime, $\Gamma_A\tau_r=0.5$, for three different values of bias voltage. The sharp peaks in the two leftmost distributions correspond to the case where the internal energy on the island is $E=2\Delta$.}
	\label{fig:currdists}
\end{figure}

The electric current through the SINIS structure near $V=2\Delta$ is given by\cite{laakso12}
\begin{equation}
 I=\frac{1}{\tau_r}\frac{\Delta}{T_t}\epsilon^{1/4}\sqrt{\pi}g_{1/2}(-u/\sqrt{\epsilon}),
\end{equation}
where $\epsilon=E/(2\Delta)$, $T_t=\sqrt{12\delta\Delta}/\pi$, $u=(V/2-\Delta)/T_t$, and $\tau_r^{-1}=gT_t^{3/2}/(\pi\sqrt{2\Delta})$. A fluctuating internal energy (or temperature) leads directly to a fluctuating electric current, and the probability distribution for the fluctuations of the electric current can be found by a change of variables from the probability distribution of energy. Three distributions in the strongly non-Gaussian regime are shown in Fig.~\ref{fig:currdists} for the dimensionless current $j=\tau_rT_tI/\Delta$. The current distributions mirror the non-Gaussian features of the distributions for internal energy on the island. For larger $u$, current increases exponentially, and the distribution turns into a Gaussian peak at larger $j$. For smaller $u$, the distribution is a narrow Gaussian peak near $j=0$.

\section{Detection of temperature fluctuations}\label{sec:cf}

\begin{figure}
    \includegraphics[width=0.7\columnwidth]{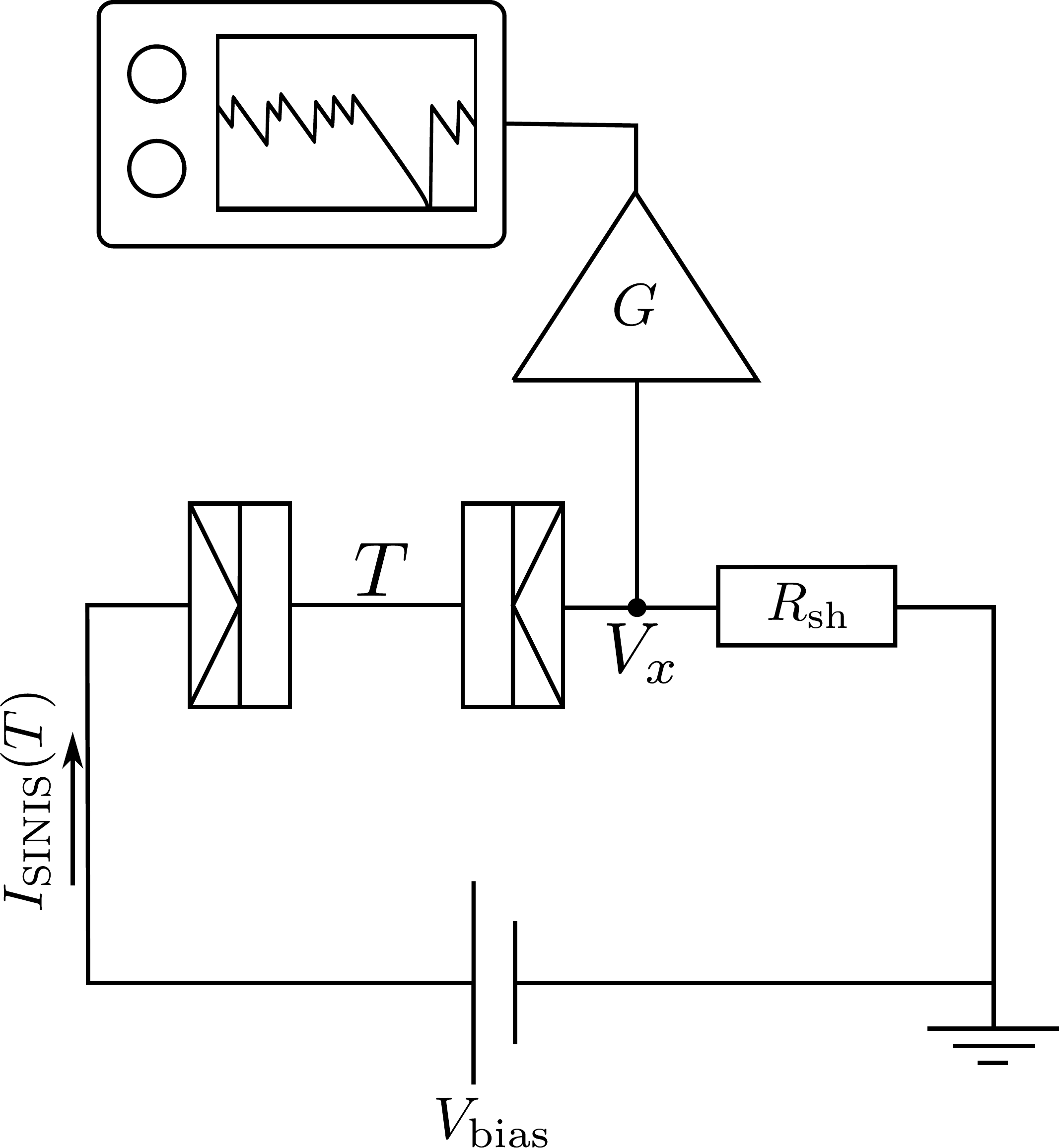}
    \caption{Example measurement scheme for the detection of temperature fluctuations via current fluctuations. The current through the SINIS structure at a fluctuating temperature $T$, $I_\mr{SINIS}(T)$, generates voltage fluctuations across the shunt resistance $R_\mr{sh}$. These are amplified and recorded in time domain.}
    \label{fig:measurement}
\end{figure}

A possible way to detect the temperature fluctuations is by detecting the induced fluctuations in some other observable. As shown in \refcite{laakso12}, the statistics of temperature fluctuations can be gathered by monitoring the current through the device in real time. The proposed measurement scheme is shown in Fig.~\ref{fig:measurement}. The temperature dependent current through the SINIS structure, $I_\mr{SINIS}(T)$, generates a fluctuating voltage $V_x$ across a shunt resistor $R_\mr{sh}$, which is then amplified and detected in time domain, with a fast oscilloscope, for example.

In order for the temperature fluctuation induced noise to be detectable, the additional noise generated by the shunt resistor and the amplifier must be smaller than the actual signal to be measured. This implies that $\sqrt{\Var(\bar{V}_x)}\ll\langle\bar{V}_x\rangle$, where the overbar denotes a time average [see Eq.~\eqref{eq:timeavg} below]. To make further progress we decompose the fluctuating voltage into a signal part and a noise part, $V_x=V+\delta V$, and average over a short measurement period $\tau\ll\tau_r$, $\tau_r$ being the relaxation time for the temperature fluctuations:
\begin{equation}\label{eq:timeavg}
    \langle\bar{V}_x\rangle=\frac{1}{\tau}\int_0^\tau\d t \langle V_x(t)\rangle=V.
\end{equation}
Variance of the measured signal is given by
\begin{align}
    \langle\bar{V}_x^2\rangle-\langle\bar{V}_x\rangle^2=&\frac{1}{\tau^2}\int_0^\tau\int_0^\tau\d t\d t' \langle\delta V(t)\delta V(t')\rangle, \nonumber \\
    =&\frac{1}{\tau^2}\int_0^\tau\int_0^\tau\d t\d t'\int\frac{\d\omega}{2\pi}e^{-\ii\omega(t-t')}S_{V_x}(\omega),
\end{align}
where $S_{V_x}(\omega)$ is the spectral noise power. Typically, the dominating source of extra noise is the thermal noise in the shunt resistor. In this case $S_{V_x}$ is independent of frequency\cite{blanter00}\footnote{Up to a frequency determined by the stray capacitances of the measurement circuit. This frequency is very high compared to the frequency scale of temperature fluctuations.} and the variance becomes
\begin{equation}
    \Var(\bar{V}_x)=\frac{S_{V_x}}{\tau}=\frac{2TR_\mr{sh}R_\mr{dyn}^2}{\tau(R_\mr{sh}+R_\mr{dyn})^2},
\end{equation}
where $R_\mr{dyn}\equiv (V_\mr{bias}-V)/\bar{I}$ is the dynamical resistance of the SINIS structure.

The signal-to-noise ratio becomes
\begin{align}
    \frac{\langle\bar{V}_x\rangle}{\sqrt{\Var(\bar{V}_x)}}=&\sqrt{\frac{R_\mr{sh}}{R_\mr{dyn}}}\sqrt{\left(\frac{V_\mr{bias}}{R_\mr{dyn}}\tau\right)\left(\frac{V_\mr{bias}}{2T}\right)}, \nonumber \\ =&\sqrt{0.58k}\left(\frac{R_\mr{sh}}{R_T}\right)^{1/2}\left(\frac{\Delta}{\delta}\right)^{1/8}\left(\frac{\Delta}{T}\right)^{1/2},
\end{align}
where on the second line we have used $\tau=k\tau_r$ and the expressions from \refcite{laakso12}. Using the numbers $k=0.1$, $R_\mr{sh}=10\ \mr{k}\Omega$, $R_T=4.3\ \mr{k}\Omega$, $\Delta=2\ \mr{K}\times k_B$, $\delta=10^{-4}\ \mr{K}\times k_B$, and $T=10\ \mr{mK}$, we get $\langle\bar{V}_x\rangle/\sqrt{\Var(\bar{V}_x)}=8$, implying that the temperature fluctuation induced noise should be readily detectable.

\section{Conclusions}\label{sec:conclusion}

Building on the recently developed theory of temperature fluctuation statistics, we have studied the temperature fluctuations and the associated current fluctuations in a SINIS structure. We have shown how the Andreev processes lead to a heating of the normal metal island and how the cooling effect near $V=2\Delta$ is accompanied by fluctuations in the island temperature. We have also shown that these temperature fluctuations lead to current fluctuations with a Fano factor that is large compared to the Fano factor of the intrinsic shot noise.

In contrast to the case of the overheated single-electron transistor,\cite{laakso10} the observation of the temperature fluctuations does not require tunnel junctions of very high resistance. The statistics of temperature fluctuations can be gathered by monitoring the electric current through the device in time domain, and extracting the statistics of temperature from the statistics of the fluctuating electric current.

In addition, the cooling effect of the SINIS structure ensures that the island temperature near $V=2\Delta$ is low enough that the electron--phonon coupling does not mask the fluctuations. Due to the poor thermal conductivity of superconductors, the practical challenge is rather to keep the superconducting reservoirs in equilibrium and at a low temperature.

\acknowledgments
We thank J.~P.~Pekola and V.~Kauppila for useful discussions. This work was supported by the Finnish Academy of Science and Letters, the Academy of Finland, and the European Research Council (Grant No.~240362-Heattronics).

\appendix
\begin{widetext}
\section{Derivation of the electron--phonon action}\label{app:derivation}

The electron--phonon action is equal to the sum of all one particle irreducible electron--phonon vacuum diagrams. The lowest order diagrams are shown in Fig.~\ref{fig:diagram}. It is enough to consider the diagram on the left, the diagram on the right vanishes due to zero momentum exchange.\cite{rammer07}
\begin{figure}
	\includegraphics[width=0.3\columnwidth]{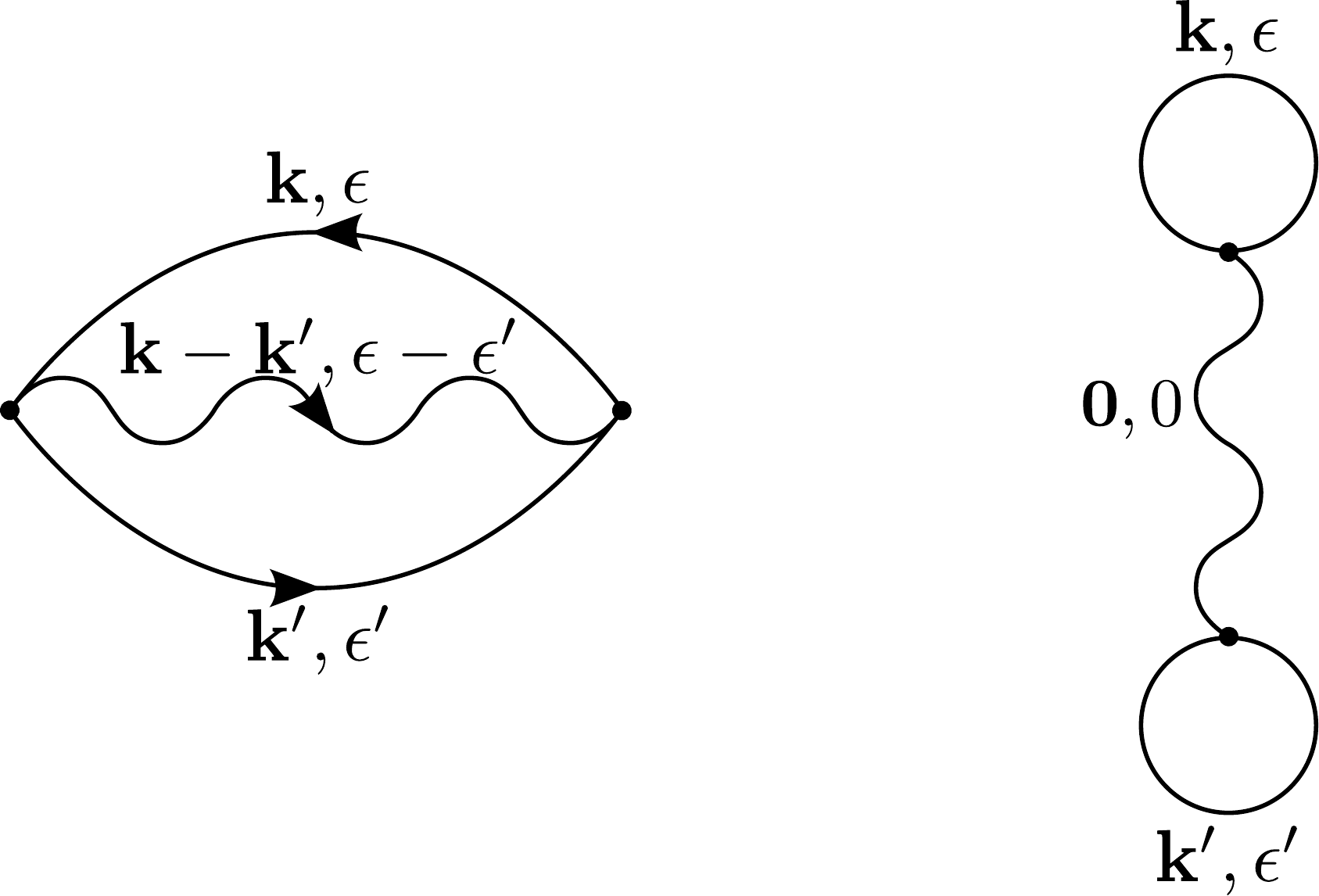}
	\caption{Two lowest order electron--phonon diagrams contributing to the effective action.}
	\label{fig:diagram}
\end{figure}
In terms of Keldysh Green's functions we have
\begin{equation}
 \mathcal{S}_{e-\mr{ph}}=\Tr \int\frac{\d\epsilon}{2\pi}\frac{\d\epsilon'}{2\pi}\frac{1}{\mathcal{V}}\sum_{\vec{k},\vec{k}'}\check{G}(\vec{k},\epsilon)\check{\gamma}|g_{\vec{k}-\vec{k}'}|^2\check{D}(\vec{k}-\vec{k}',\epsilon-\epsilon')\check{G}(\vec{k}',\epsilon')\check{\tilde{\gamma}},
\end{equation}
where $\Tr$ is trace over Keldysh and Nambu indices. The absorption and emission vertices are, respectively,\cite{rammer07}
\begin{equation}
 \check{\gamma}^k_{ij}=\delta_{ij}\delta_{jk},\quad\check{\tilde{\gamma}}^k_{ij}=\delta_{ij}\tau^{(3)}_{jk}.
\end{equation}
Since the electrons participating in the interactions lie close to the Fermi surface, we can use the quasiclassical approximation to write
\begin{equation}
 \mathcal{S}_{e-\mr{ph}}=\Tr \int\frac{\d\epsilon}{2\pi}\frac{\d\epsilon'}{2\pi}\frac{\mathcal{V}\nu^2}{16}\int\d S_F\d S'_F\check{G}(\vec{e}_\vec{k},\epsilon)\check{\gamma}|g_{k_F(\vec{e}_\vec{k}-\vec{e}_{\vec{k}'})}|^2\check{D}(k_F(\vec{e}_\vec{k}-\vec{e}_{\vec{k}'}),\epsilon-\epsilon')\check{G}(\vec{e}_{\vec{k}'},\epsilon')\check{\tilde{\gamma}},\end{equation}
where $\d S_F$ is an area element of the Fermi surface, $\nu$ the density of states at the Fermi surface, and $\vec{e}_\vec{k}$ unit vector in the direction of $\vec{k}$.

Assuming that phonons are not affected by the electrons, i.e., that they are in equilibrium at temperature $T_\mr{ph}$, $D^K=(D^R-D^A)\coth(\epsilon/2T_\mr{ph})$. For the electron Green's function we use Eq.~\eqref{eq:green}. Carrying out the matrix and tensor operations results in
\begin{align}\label{eq:ephactionnambu}
 \mathcal{S}_{e-\mr{ph}}=&\int\frac{\d\epsilon}{2\pi}\frac{\d\epsilon'}{2\pi}\frac{\mathcal{V}\nu^2}{16}\int\d S_F\d S'_F|g_{k_F(\vec{e}_\vec{k}-\vec{e}_{\vec{k}'})}|^2(D^R-D^A)\Bigl\{4N(\epsilon)N(\epsilon')\Bigr. \nonumber \\ &\times\left.\left[e^{\xi(\epsilon-\epsilon')}n(\epsilon-\epsilon')\left(f(\epsilon')(1-f(\epsilon))+(1-f(-\epsilon'))f(-\epsilon)\right)\right.\right. \nonumber \\ & \left.+e^{-\xi(\epsilon-\epsilon')}\left(1+n(\epsilon-\epsilon')\right)\left(f(\epsilon)(1-f(\epsilon'))+(1-f(-\epsilon))f(-\epsilon')\right)\right] \nonumber \\
 &-e^{\xi(\epsilon-\epsilon')}n(\epsilon-\epsilon')\left[\left(f^R(\epsilon)(1-f(\epsilon))+f^A(\epsilon)f(-\epsilon)\right)\left(f^R(\epsilon')(1-f(-\epsilon'))+f^A(\epsilon')f(\epsilon')\right)\right. \nonumber \\ &+\left.\left(f^R(\epsilon)f(-\epsilon)+f^A(\epsilon)(1-f(\epsilon))\right)\left(f^R(\epsilon')f(\epsilon')+f^A(\epsilon')(1-f(-\epsilon'))\right)\right] \nonumber \\ &-e^{-\xi(\epsilon-\epsilon')}\left(1+n(\epsilon-\epsilon')\right)\left[\left(f^R(\epsilon)f(\epsilon)+f^A(\epsilon)(1-f(-\epsilon))\right)\left(f^R(\epsilon')f(-\epsilon')+f^A(\epsilon')(1-f(\epsilon'))\right)\right. \nonumber \\ &+\Bigl.\left.\left(f^R(\epsilon)(1-f(-\epsilon))+f^A(\epsilon)f(\epsilon)\right)\left(f^R(\epsilon')(1-f(\epsilon'))+f^A(\epsilon')f(-\epsilon')\right)\right]\Bigr\}
\end{align}
where 
\begin{equation*}
N(\epsilon)=\mr{Re}\left(\frac{|\epsilon|}{\sqrt{\epsilon^2-\Delta^2}}\right),\quad f^{R(A)}(\epsilon)=\frac{\Delta}{\sqrt{(\epsilon\pm\ii0^+)^2-\Delta^2}}\sgn(\epsilon).
\end{equation*}
If $f(-\epsilon)=1-f(\epsilon)$ this simplifies to
\begin{align}\label{eq:ephactionnambu2}
 \mathcal{S}_{e-\mr{ph}}=&\int\frac{\d\epsilon}{2\pi}\frac{\d\epsilon'}{2\pi}\frac{\mathcal{V}\nu^2}{2}\int\d S_F\d S'_F|g_{k_F(\vec{e}_\vec{k}-\vec{e}_{\vec{k}'})}|^2(D^R-D^A)\left(N(\epsilon)N(\epsilon')-F(\epsilon)F(\epsilon')\right) \nonumber \\ &\times\left(e^{\xi(\epsilon-\epsilon')}n(\epsilon-\epsilon')f(\epsilon')(1-f(\epsilon))+e^{-\xi(\epsilon-\epsilon')}\left(1+n(\epsilon-\epsilon')\right)f(\epsilon)(1-f(\epsilon'))\right),
\end{align}
where
\begin{equation*}
F(\epsilon)=\mr{Re}\left(\frac{\Delta\sgn(\epsilon)}{\sqrt{\epsilon^2-\Delta^2}}\right).
\end{equation*}

Let us now concentrate on the spectral function of phonons. In equilibrium
\begin{equation}
 A(\vec{q},\epsilon)=\ii\left(D^R(\vec{q},\epsilon)-D^A(\vec{q},\epsilon)\right)=2\pi\left(\delta(\epsilon-\varepsilon_\vec{q})-\delta(\epsilon+\varepsilon_\vec{q})\right),
\end{equation}
where $\varepsilon_\vec{q}=v_D|\vec{q}|$ is the (acoustical) phonon dispersion relation with the speed of sound $v_D$. In the jellium model the e-ph coupling is
$|g_\vec{q}|^2=\tilde{g}^2\varepsilon_\vec{q}/2$, where $\tilde{g}$ is now some dimensionless coupling constant. For a spherical Fermi surface, and any function $f$,
\begin{equation}
 \int\d S_F\d S'_F f(|\vec{e}_\vec{k}-\vec{e}_{\vec{k}'}|)=8\pi^2\int_0^2\d k kf(k).
\end{equation}
Using this result we get
\begin{equation}
 \int\d S_F\d S'_F|g_{k_F(\vec{e}_\vec{k}-\vec{e}_{\vec{k}'})}|^2A(k_F(\vec{e}_\vec{k}-\vec{e}_{\vec{k}'}),\epsilon-\epsilon')=\frac{8\pi^3\tilde{g}^2}{\varepsilon^2_\mr{max}}(\epsilon-\epsilon')^2\sgn(\epsilon-\epsilon')\theta(2\varepsilon_\mr{max}-|\epsilon-\epsilon'|),
\end{equation}
where $\varepsilon_\mr{max}=v_Dk_F$. This can now be inserted to Eq.~\eqref{eq:ephactionnambu2}, which after identifying $\Sigma=24\pi\tilde{g}^2\nu^2\zeta(5)/\varepsilon^2_\mr{max}$ leads to Eq.~\eqref{eq:e-ph} for a normal metal island.

\end{widetext}

\bibliography{fcs}

\end{document}